\documentstyle[prl,aps,twocolumn,epsf]{revtex}
\begin{document}

\twocolumn[
\hsize\textwidth\columnwidth\hsize\csname @twocolumnfalse\endcsname

\title{Charge-density-wave formation by Van Hove nesting in
the $\alpha$-phase of Sn/Ge(111)}
\author{J. Gonz\'alez \\}
\address{
        Instituto de Estructura de la Materia.  Consejo Superior
de Investigaciones Cient{\'\i}ficas.  Serrano 123, 28006 Madrid.
Spain.}
\date{\today}
\maketitle
\begin{abstract}
\widetext
We study the role of electron correlations
in the formation of the surface charge-density-wave state
in the Sn/Ge(111) interface. The Fermi energy of the
overlayer is treated as a dynamical variable, which
undergoes a substantial renormalization by the interaction. 
We show that the Fermi
level turns out to be pinned to a Van Hove singularity in the density of
states, which explains the formation of the
charge-density-wave, the observation of a very flat band
in photoemission experiments and the reduction of the spectral
weight in the low-temperature phase.
\end{abstract}
\pacs{73.20.At,71.27.+a,71.45.Lr}

]

\narrowtext 
\tightenlines
The formation of states with rearrangement of the electronic charge 
is accompanied in general by interesting properties in condensed
matter systems. A recent example has been
provided by the study of interfaces like 
Pb/Ge(111)\cite{carp,gold},
Sn/Ge(111)\cite{carpi,modes} or K/Si(111)\cite{weit}. 
All of them have in common the
existence of a phase transition with the formation of a surface
charge-density-wave (CDW) in the low-temperature phase. 
In the first two cases
the adsorbate phases have at room temperature the same 
$(\sqrt{3} \times \sqrt{3})R30^{\circ} $ structure, and they
both show a reduction of the symmetry to a $(3 \times 3)$ ground
state below the critical temperature. However,
while the Pb/Ge(111) overlayer becomes
insulating in the charge-ordered state, the Sn/Ge(111) remains
metallic below the transition. In the first case, the analysis
of the surface band has revealed nesting along significant
portions of the two-dimensional Fermi surface\cite{carp}, 
which would account for the formation of the CDW. In the 
$\alpha$-phase of Sn/Ge(111), however, no similar effect seems
to be at work\cite{carpi,modes}, 
and the reason why the ground state of the system is led to
reorder its charge is not fully understood yet.

It is believed that correlation effects may play an important role 
in the development of the charge-ordered state\cite{tos1,tos2}. 
The relatively
small surface bandwidth of these systems ($\sim 0.5$ eV) leads to
consider the on-site Coulomb repulsion as one of the relevant
interactions. It has been also suggested that a nearest-neighbor
repulsion may play an important role in producing the $3 \times
3$ surface periodicity\cite{tos1}.

In this letter we develop the idea that the electron correlations 
are essential to understand the low-temperature phase of 
Sn/Ge(111), while relying on the fact that no strong-coupling 
effects need to be invoked for that purpose. This is plausible 
as long as the overlayer remains metallic below the transition. 
We will show that the 
$(\sqrt{3} \times \sqrt{3})$ phase is unstable towards the
formation of saddle points near the Fermi surface in the
electronic dispersion relation. This is in agreement with
the observation of a very flat band below the Fermi level
in the angle-resolved photoemission experiments carried out
in the CDW phase\cite{modes}. 
The Van Hove singularity
controls the dynamics of the system since, as we will see, this
finds energetically favorable to readjust the bands in order to
place the Fermi surface closer to the saddle points. As well as
nesting of the saddle points near the Fermi surface gives rise to
a spin-density-wave instability in the repulsive $t-t'$ Hubbard 
model\cite{mark,ioffe,alvarez,murakami,metz},
it is able to explain the formation of the CDW 
under the assumption of a more general interaction in the
present instance.

We will take an extended Hubbard model with on-site $U$ and
nearest-neighbor $V$ repulsive interactions on the triangular lattice
of the Sn/Ge(111) overlayer to illustrate these ideas. The
adatom dangling bonds give rise to a band that reaches a minimum
at the corner of the hexagonal $(\sqrt{3} \times \sqrt{3})$
Brillouin zone, and a maximum close to the center\cite{carpi}.
We use a tight-binding approximation with nearest neighbor
hopping $t$ and next-to-nearest neighbor hopping $t'$ to
parametrize the Fermi line of the $(\sqrt{3} \times \sqrt{3})$
phase as given in Ref. \onlinecite{carpi}.
Although the precision of the fit to the dispersion relation is 
limited with this approximation, we are able to capture in this way
the main topological features, which play an essential role in
the following considerations.

\begin{figure}
\begin{center}
\epsfysize=6cm
\mbox{\epsfbox{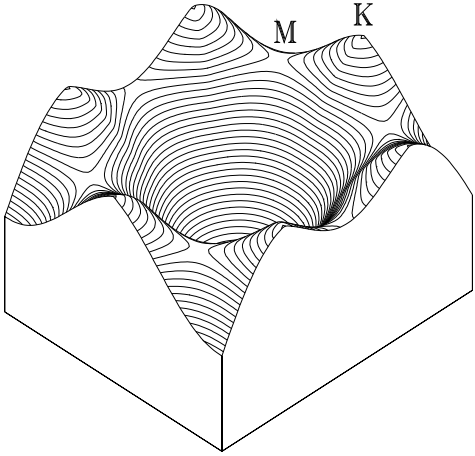}}
\end{center}
\caption{Lower subband of the $(3 \times 3)$ phase. The upper subband
(not shown) displays symmetric cusps at the six corners of the
$(3 \times 3)$ Brillouin zone.}
\label{one}   
\end{figure}

\begin{figure}
\begin{center}
\epsfysize=6cm 
\mbox{\epsfbox{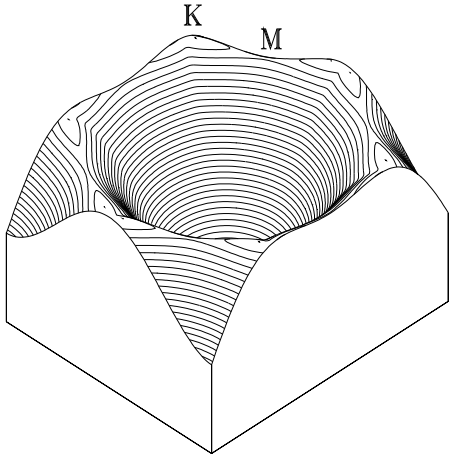}}
\end{center}
\caption{Middle subband of the $(3 \times 3)$ phase. The saddle
points lie at the boundary of the $(3 \times 3)$ Brillouin zone.}
\label{two}
\end{figure}

The topology of the dispersion relation is actually unstable under 
the smallest periodic lattice distortion. 
This has to be taken into account,
as it has been recently observed that one of the Sn atoms in 
the $(3 \times 3)$ unit cell is slightly displaced upwards 
with respect to the other two\cite{mca,bunk}. 
By taking a hopping parameter equal to $0.9 t$ 
between inequivalent Sn atoms in the $(3 \times 3)$ unit cell,
we already observe the appearance of two subbands
of very different character. The original valence band for the
$(\sqrt{3} \times \sqrt{3})$ lattice splits along the 
$\Gamma$K direction of the $(3 \times 3)$ Brillouin zone.
Thus, there are an upper and a lower subband that join at the apex of
the conical dispersion shown in Fig. \ref{one}, 
characteristic of the underlying honeycomb lattice, 
and a  middle subband that develops a saddle point at
the boundary of the Brillouin zone, as shown in Fig. \ref{two}. 
The density of states redistributes itself,
having a linear dependence in energy near K in the
upper and lower honeycomb-type subbands while becoming divergent at 
the M point in the middle subband.

Before switching the interaction, the Fermi level has a 
nominal location below the mentioned Van Hove singularity.
However, the relative position of the Fermi level in the 
different subbands of the model may suffer significant changes by
the effect of the electron correlations near the singularity.
The forward scattering vertex $\Gamma_{\rm fwd}$ giving the four-point
interaction at small momentum transfer ${\bf k}$ gets corrected by the
iteration of the particle-hole polarizability. For the saddle-point
dispersion relation $\varepsilon ({\bf k}) \approx k_x^2 - k_y^2 $, 
this object diverges in the static limit as 
$\sim \log (\Lambda  /{\rm max} \{ |\varepsilon ({\bf k})| , 
 |\varepsilon_F | \} )$, where $\varepsilon_F $ is the distance of
the Fermi level to the 
Van Hove singularity and $\Lambda $ is a cutoff of the order of the 
bandwidth\cite{nucl}. 
In the case of a
spin-independent interaction, this leads to a reduction of the
vertex at zero momentum-transfer 
\begin{equation}
\Gamma_{\rm fwd} (\varepsilon_F) = \frac{U_{\rm fwd}}
      {1 + (c/(2\pi^2 \Lambda))U_{\rm fwd} \log | \Lambda /\varepsilon_F |}
\label{renorm}
\end{equation}
where the bare coupling is $U_{\rm fwd} \equiv U({\bf p}_1 , 
{\bf p}_2 ; {\bf p}_1 , {\bf p}_2) - U({\bf p}_1 ,
{\bf p}_2 ; {\bf p}_2 , {\bf p}_1)$, in terms of the original 
four-fermion interaction, and 
$c$ is the prefactor of the particle-hole polarizability\cite{us}.
This renormalization of $\Gamma_{\rm fwd}$ has important consequences,
since it encodes the sum of dominant logarithmic contributions
that give rise to the constant shift of the chemical potential in the 
self-energy corrections\cite{sub}. 
In this respect, it has
been shown that the Van Hove singularity has a natural tendency
to pin the Fermi level in a model with bare repulsive
interactions\cite{mark2,euro}. 
In the present case, this can be understood in
terms of the uneven screening of the forward scattering
in the different subbands of the model, as we show in what
follows.

Let us model the problem of the interaction between the two 
subbands in Figs. \ref{one} and \ref{two} by taking the divergent 
density of states near the saddle-point structure 
\begin{equation}
n^{(1)} (\varepsilon) = - \frac{1}{\Lambda } \log (|\varepsilon |  
  /\Lambda ) \;  ,  -\Lambda < \varepsilon < \Lambda
\label{n1}
\end{equation}
and a vanishing density of states for the lower subband
\begin{equation}
n^{(2)} (\varepsilon) =  \frac{\alpha}{\Lambda^2 } |\varepsilon |
 \;  ,  - \beta \Lambda < \varepsilon < \beta \Lambda
\label{n2}
\end{equation}

The physical value of the Fermi energy comes from the balance
between the nominal value $\mu $ of the chemical potential and
the shift of the one-particle levels to higher energies due to the
repulsive interaction.
The filling level of each subband is determined by the self-energy
corrections, bearing in mind that the upward displacement of the levels
depends in turn on the charge present in all the subbands. 
The filling level $\varepsilon_{F1}$ 
in the middle subband is given by the equation
\begin{equation}
\varepsilon_{F1} = \mu - \Gamma_{\rm fwd} (\varepsilon_{F1}) 
  \int^{\varepsilon_{F1}} d \varepsilon n^{(1)} (\varepsilon )
  - g_{\rm fwd} \int^{\varepsilon_{F2}} d \varepsilon n^{(2)} 
      (\varepsilon )
\label{self1}
\end{equation}
where we have introduced a coupling constant $g_{\rm fwd}$ that
parametrizes the Coulomb repulsion exerted on the middle subband 
by the charge present in the lower subband.
Analogously, the filling level $\varepsilon_{F2}$ in the latter
is computed in the form
\begin{equation}
\varepsilon_{F2} = \mu - U_{\rm fwd} 
  \int^{\varepsilon_{F2}} d \varepsilon n^{(2)} (\varepsilon )
  - g_{\rm fwd} \int^{\varepsilon_{F1}} d \varepsilon n^{(1)} 
      (\varepsilon )
\label{self2}
\end{equation}

We remark that $\varepsilon_{F1}$ and
$\varepsilon_{F2}$ are measured in the reference frames in which 
the dependences $n^{(1)} (\varepsilon) $ and 
$n^{(2)} (\varepsilon)$ are as given in Eqs. (\ref{n1}) and
(\ref{n2}). Thus, the
fact that $\varepsilon_{F1}$ and
$\varepsilon_{F2}$ are nominally different after renormalization
is just an artifact of that convention. The physical picture
is however the opposite, namely that the one-particle levels are 
shifted to higher energy by a different amount in each of the 
subbands, up to a point in which the
respective Fermi levels reach the common chemical potential.

\begin{figure}  
\begin{center}
\epsfysize=7cm
\mbox{\epsfbox{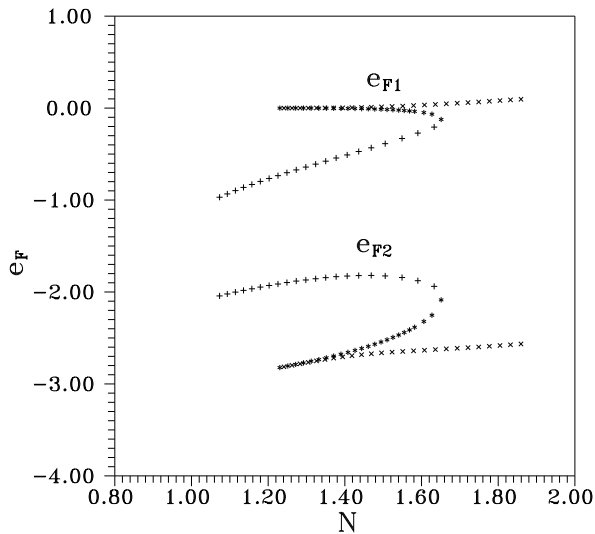}}
\end{center}
\caption{Self-consistent solutions for the filling
levels in the middle and lower subbands vs total charge $N$ in
the overlayer.}
\label{three} 
\end{figure}

The coupled set of equations (\ref{self1}) and (\ref{self2})
gives rise to nontrivial physical effects, as a consequence of
the nonlinearities introduced by the divergent density of states
$n^{(1)} (\varepsilon) $ and the renormalization of
$\Gamma_{\rm fwd} (\varepsilon)$ close to the Van Hove singularity.
It is convenient to solve for the location of
$\varepsilon_{F1} $ and $\varepsilon_{F2} $ in terms of the
total charge $N$ in the system, given by
\begin{equation}
N = \int^{\varepsilon_{F1}} d \varepsilon n^{(1)} (\varepsilon )
  + \int^{\varepsilon_{F2}} d \varepsilon n^{(2)} (\varepsilon )
\end{equation}
The most remarkable effect is that there is not a one-to-one
correspondence between $N$ and the respective filling levels 
$\varepsilon_{F1}$ and $\varepsilon_{F2}$ for the two subbands.
The different branches of the solution are represented in 
Fig. \ref{three} for the particular values $U_{\rm fwd} 
= 4.0 \Lambda $ and $g_{\rm fwd} = 3.0 \Lambda $. The parameter
$\beta $ has been chosen equal to $3.0$, and $\alpha $ has been
set so that there is the same number of states in the lower and
in the middle subband.

\begin{figure}
\begin{center}
\epsfysize=7cm
\mbox{\epsfbox{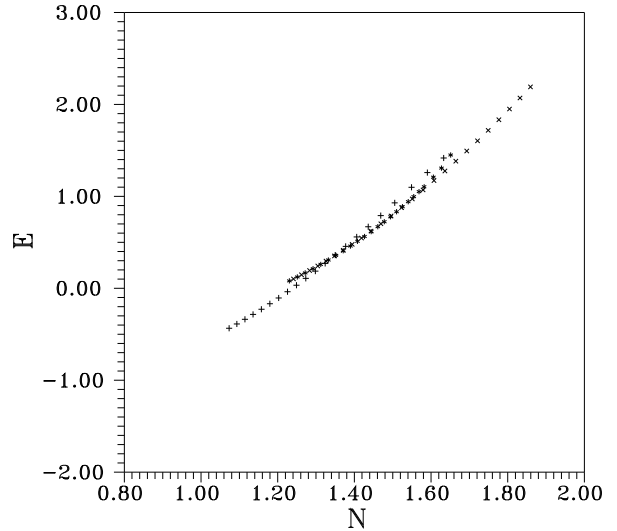}}
\end{center}
\caption{Total energy of the different solutions
shown in Fig. \ref{three} vs total charge $N$ (the symbols for
a given solution match in the two figures).}
\label{four}
\end{figure}

At low values of $N$, the filling of the middle subband with the
Van Hove singularity proceeds in a regular way, with a
monotonous increase of $\varepsilon_{F1} $. There is a point,
however, in which two other locations of $\varepsilon_{F1}$
become possible, closer to the singularity in the density of
states. In these instances, the corresponding filling level
$\varepsilon_{F2} $ in the second subband suffers an abrupt
decrease with respect to the expected value. It is interesting
to discern what of the possible solutions is most favorable
energetically. We have plotted in Fig. \ref{four} the value of
the total energy $E$ versus the total charge $N$. We see that
above a certain value, that is $N \approx 1.4 $ in our case,
any of the two filling levels close to the Van Hove singularity
give the most stable configuration of the system. This is
in agreement with previous analyses of the pinning of the Fermi 
level of electrons near a Van Hove singularity\cite{mark2,euro}.

The result that turns out to be valid under very general
conditions is the existence of an intermediate range of filling
levels that are forbidden below the Van Hove singularity. For
the corresponding values of the charge $N$, this finds more
favorable to fill the Fermi sea up to the Van Hove singularity,
at the expense of the charge in the other subband. In general,
there is a critical value of $N$ in which the filling level 
$\varepsilon_{F1}$ jumps discontinously from the regular
evolution upon doping to a position much closer to the Van Hove
singularity. 

Coming back to the physics of the Sn/Ge(111)
overlayer, a description consistent with the plots in Figs.
\ref{one} and \ref{two} leads to a nominal value of 
$\varepsilon_{F1} $ that
is about $25 \% $ off the Van Hove singularity, relative to the
energy difference between the level of the latter and the bottom
of the subband. 
This nominal value falls certainly in the forbidden region,
what leads us to conclude according to the above arguments that
the Fermi level is pinned close to the Van Hove singularity
developed in the system. 

This abrupt renormalization of the Fermi level towards the Van
Hove singularity is able to account for all the prominent
experimental features found in the Sn/Ge(111) interface, as we
show in what follows.

\vspace{0.3cm}

\noindent
1. Strong correlations in the low-temperature phase.
 
The different photoemission
experiments have in common the observation in the
low-temperature phase of a lower band that disperses along the
$\Gamma$M direction of the $(3 \times 3)$ Brillouin
zone, together with another band with higher energy and
much less dispersion\cite{modes,mca}. 
The point is that this intermediate band is
found at an energy sensibly lower than predicted by LDA
calculations\cite{mca}. 
Quite remarkably, all the points along the $\Gamma$M
direction are found below the Fermi level. 
After taking into account the experimental overestimation
of the binding energies remarked in Ref. \onlinecite{modes},
the M point turns out to be close but {\em below} the 
Fermi level, contrary to the estimates
carried out neglecting many-body effects. 
This is actually the
experimental signature of the pinning to the Van Hove singularity 
we have discussed above.

It is also clear that the width of the middle subband is sensibly 
smaller than predicted by LDA calculations\cite{modes,mca}. 
This reduction in the bandwidth is also
consistent with the renormalization of the kinetic energy of
electrons close to a saddle point, which leads to a flatter 
shape of the dispersion relation\cite{nucl}. 

\vspace{0.3cm}

\noindent
2. Formation of the surface CDW.

The saddle points of the middle subband are at the M points 
of the $(3 \times 3)$
Brillouin zone, so that they are separated by momenta ${\bf
K}_i$ that correspond actually to the wavevectors of the CDW
structure observed experimentally. This Van Hove nesting by momenta 
${\bf K}_i$ connecting the saddle points near the Fermi level leads 
to the CDW instability.

In the context of an extended Hubbard model and with the
inclusion of the electron-phonon interaction,
it has been shown that it is natural to have a negative
coupling $U_{{\bf K}_i}$ for the four-fermion interaction
with momentum transfer ${\bf K}_i$\cite{tos1}
(its bare value is actually $U - 6V$, in terms of the
parameters of the extended Hubbard model).
We remark that, with a divergent density of 
states close to the Fermi level, the slightest attractive interaction
($U_{{\bf K}_i} < 0$) triggers the CDW instability.  
The particle-hole susceptibility with momentum ${\bf K}_i$ is
itself logarithmically divergent, and the iteration of 
particle-hole diagrams leads to an effective vertex at momentum
transfer ${\bf K}_i$
\begin{equation}
\Gamma_{{\bf K}_i} (\varepsilon ) = \frac{U_{{\bf K}_i}}
{1 + (c'/(2\pi^2 \Lambda))U_{{\bf K}_i} \log | \Lambda
       /\varepsilon |}
\label{eff}
\end{equation}
Thus, the attractive interaction is overscreened and renormalized
towards more intense attraction at low energies.


Under these conditions, the only response function that
shows a divergence at some critical frequency corresponds to 
a charge modulation with wavevectors ${\bf K}_i$. 
The gap that opens up at the saddle points can be actually
estimated from the frequency at which the denominator in Eq. 
(\ref{eff}) vanishes. The order of magnitude thus obtained
(tens of meV) is consistent with the temperature of the 
transition to the CDW.

\vspace{0.3cm}

\noindent
3. Metallic behavior and depletion of the density of states in
the low-temperature phase.

The metallic behavior of the $(3
\times 3)$ phase follows naturally from our description of the
different subbands in the system. Nesting of the saddle points 
at the Van Hove filling and the formation of the
surface CDW lead to the opening of a gap and to a partial
destruction of the Fermi line around the saddle points\cite{rice}.
The divergent density of states induces by itself a significant
reduction of the quasiparticle weight, as appreciated in the
computation of the self-energy corrections for the saddle point
dispersion relation\cite{nucl}. 
This is consistent with the decrease of the
photoemission intensity about the Fermi level observed in the
CDW phase\cite{modes}. The lower band is less sensitive to these
renormalization effects, and the metallic properties come mainly
from the band crossing the Fermi level below the cusps shown in
Fig. \ref{one}.

\vspace{0.3cm}

In conclusion, the pinning of the
Fermi level to a Van Hove singularity in the density of states
leads to a phenomenology for the Sn/Ge(111) overlayer
consistent with the experimental
observations of CDW formation and photoemission measurements,
which are challenging more conventional theoretical approaches.
Further
predictions of our approach should also be tested experimentally
in Sn/Ge(111), in particular the renormalization of the lower 
band and the geometry of the Fermi line around the corners 
of the $(3 \times 3)$ Brillouin zone.


\begin{references}
\bibitem{sub}
Momentum dependence is carried however by the subdominant terms, which
are responsible of a mild correction of the shape of the
saddle-point dispersion relation.
\end{references}
\end{document}